\documentclass{svjour3}                     
\smartqed  
\usepackage{graphicx}
\journalname{myjournal}
\begin{document}

\title{Continuous matter creation and the acceleration of the universe:
the growth of density fluctuations}

\titlerunning{growth of density fluctuations}        

\author{Alain de Roany  \and J.A. de Freitas Pacheco}


\institute{Universit\'e de Nice-Sophia Antipolis-Observatoire de la C\^ote d'Azur\\
BP 4229, 06304, Nice Cedex 4, France\\
\email{deroany@oca.eu, pacheco@oca.eu}           
}

\date{Received: date / Accepted: date}

\maketitle

\begin{abstract}

Cosmologies including continuous matter creation are able to reproduce
the main properties of the standard $\Lambda$CDM model, in particular in cases 
where the particle and entropy production rates are equal. These specific models,
characterized by a mass density equal to the critical value, 
behave like the standard $\Lambda$CDM model at early times whereas their late evolution is similar
to the steady-state cosmology. The maximum amplitude of density fluctuations in 
these models depends on the adopted creation rate, related here to the parameter $\Omega_v$ and 
this limitation could be a difficulty
for the formation of galaxies and large-scale structure in this class of universe.
Additional problems are related with predictions either of the random peculiar velocities of galaxies
or the present density of massive clusters of galaxies, both being largely overestimated
with respect to observational data.

\end{abstract}

\section{Introduction}

The standard model in cosmology, the so-called $\Lambda$CDM model, assumes 
the presence of a constant cosmological term in Einstein's equations and that
the universe is constituted, besides baryons, photons and neutrinos, by 
a dominant weakly interacting component of unknown nature dubbed dark matter. This model gives the 
best description of the present data as, for instance, those resulting from the 
analysis of the seven years operation of the WMAP satellite \cite{wmap} and supernova distances \cite{union}.

The inclusion of the cosmological term in Einstein's equations represents the most simple 
and natural explanation for the observed acceleration of the expansion of the universe 
detected through supernova data. However, arguments against the inclusion of such a
cosmological term are often found in the literature. The first one is the so-called ``coincidence" problem, which
can be stated in the following way: why do we observe presently an almost equal contribution
of matter and the cosmological term to the total energy budget of the universe, considering
that these components evolve differently with time? Such a ``coincidence" would suggest that
we are living in a particular moment of the history of the universe, contrary to a ``cosmological
principle" stating that we are not in a special place in the universe either in space or in time.
Notice that if the accelerated phase started around $z \approx 0.75$, such 
a phase covers about 48\% of the existence of our universe!
The second argument is related to the interpretation of the cosmological constant as the
vacuum energy density. As it is well known, formal quantization of classical fields leads to a
divergent energy density for the vacuum state, which can be avoided by imposing a physical
cut-off \cite{mukha}. If we adopt as a cut-off the Planck scale, then the theory predicts an effective
cosmological constant corresponding to an energy density that is about 120 orders of magnitude 
higher than the observed
value. Convincing arguments against these difficulties were recently given by Bianchi
and Rovelli \cite{rovel}, who concluded that the ``coincidence problem" is ill defined and that the identification
of the cosmological constant with the vacuum energy density is probably a mistake.

Face to this debate on the existence (or not) of the cosmological term, alternative models 
have been discussed in the literature \cite{germano}, \cite{zimdahl}, \cite{triginer}, \cite{limaalcaniz},
\cite{alcaniz}. In some of these models, the acceleration of the expansion is driven by a negative pressure 
term associated to particle creation at the expense of the gravitational field, an original idea proposed 
by Zeldovich \cite{zel} almost 40 years ago. Particular cases in which the particle production rate is
proportional to the Hubble parameter give results similar to the 
canonical $\Lambda$CDM model \cite{card}, \cite{lim}. 
From a thermodynamic point of view, particle production from the gravitational field can be considered
in the context of open systems, where the ``heat" received by the system is due entirely to the change in the
number of particles \cite{prig}. In the standard model, matter is produced suddendly at the initial
singularity and during the re-heating at the end of the inflationary phase. After that the universe evolves
adiabatically, i.e., both the entropy per unit of comoving volume and the entropy per particle
remain constant during the expansion.

In the present paper we review the properties of specific models including particle creation, emphasizing some
particular aspects not previously noticed. We investigate also the linear growth of density 
perturbations in this class of
cosmological models and we will show that, depending on the creation rate, the amplitude of the density
contrast is unable to grow beyond a certain limit. The critical creation rate below which density
fluctuations are able to grow continuously up to present time is marginally compatible
with the parameters of the $\Lambda$CDM moldel. This could represent a difficulty for the formation of
galaxies and the large-scale structure, condemning this particular class of cosmology. This paper 
is organized as follows: in Section 2 the principal
physical aspects of the model are revisited; in Section 3 the growth of density perturbations is
examined and finally, in Section 4 the main conclusions are given.

\section{The cosmological model}

In the present investigation, for simplicity, we will neglect the evolution of the baryonic 
component and we will focus our attention on the dominant (dark matter) component only.
The number of baryons per unit of comoving volume remains conserved during the expansion, i.e., 
$n_ba^3$ = constant, where $n_b$ is the comoving baryon number density and
$a=a(t)$ is the scale factor. On the other hand, the bulk of the dominant component is produced as
in the standard scenario but now the possibility of a continuous creation
of dark matter particles is included. This implies that we renounce to the common 
idea that the expansion of the universe
could be described by ``closed" system and we assume, following Prigogine et al. \cite{prig} that, in fact,
the expansion is described by an ``open" thermodynamic system. Under these conditions,
the first law can be written as
\begin{equation}
TdS = dE+PdV-\mu dN
\label{firstlaw}
\end{equation}
where the chemical potential $\mu$ is here associated to the non conservation of the particle number. Introducing
respectively the entropy density $s$, the energy density $\rho$, the particle number density $n$, the
enthalpy density $h=(P+\rho)$, replacing into eq.~\ref{firstlaw}, developing and using the Euler's relation
\begin{equation}
n\mu = (P+\rho)-Ts = h-Ts
\label{euler}
\end{equation}
one obtains after some algebra
\begin{equation}
\left(\frac{d\rho}{dt}-\frac{h}{n}\frac{dn}{dt}\right)=sT\left(\frac{d\lg s}{dt}-\frac{d\lg n}{dt}\right)
\label{gibbs}
\end{equation}
which is essentially the Gibbs relation for the system (see \cite{cal} for a covariant derivation
of this relation).

Solutions of eq.~\ref{gibbs} permit the study of different particle production scenarios but relations
either between the state variables $s$ and $n$ or for their production rates are required, since they
define the model. The simplest
possibility, considered either by \cite{card} or \cite{lim} , corresponds to the case where the entropy per particle
remains constant during the expansion process, i.e., $s/n = constant$. This hypothesis implies 
that the energy density a function of the particle number density only
and not of the temperature also. Moreover, it is worth mentioning that such
a condition is not equivalent to an adiabatic expansion as it occurs in the standard model since, as we shall
see later, there is entropy production and the aforementioned condition expresses only the fact
that the relative rates of particle and entropy production are equal.
In this case, the right side 
of eq.~\ref{gibbs} is zero and the particle production rate is related to the energy density variation rate
by
\begin{equation}
\frac{d\rho}{dt}=\frac{h}{n}\frac{dn}{dt}
\label{rate}
\end{equation}
Let the stress-energy tensor of the dark matter fluid be
\begin{equation}
T_{ab}= (\rho + \Pi)u_au_b -\Pi g_{ab}
\label{stress}
\end{equation}
where $\Pi = P(\rho)+P_c$ is the effective pressure acting on the fluid, with the
first term on the right side representing the pressure due to kinetic motions and interactions between
particles and the second, $P_c$, being a new term associated to the particle production. Considering 
a Friedman-Robertson-Walker(FRW) metric for a spatially flat spacetime, i.e.,
\begin{equation}
ds^2 = -dt^2 + a^2(t)\left(dr^2+r^2d\Omega^2\right)
\label{metric}
\end{equation}
from the equation $T^k_{0;k}=0$, one obtains for the variation rate of the energy density
\begin{equation}
\frac{d\rho}{dt}+3\frac{\dot a}{a}\left(h+P_c\right)=0
\label{energy}
\end{equation}
Then, from eqs.~\ref{rate} and \ref{energy}, one obtains immediately an expression for the pressure associated to
particle production
\begin{equation}
P_c = -\frac{h}{3H}\left(3H+\frac{d\lg n}{dt}\right)
\label{particlerate}
\end{equation}
where the Hubble parameter $H$ was defined as usually, namely, $H=\dot a/a$. It should be emphasized again
that the above relation
is valid only if the ratio $s/n$ remains constant in the process. As expected, the 
relation above indicates that if the particle production rate is positive in 
a expanding universe ($H > 0$), the pressure $P_c$ is negative, contributing to accelerate the expansion.
The simplest world model without a cosmological constant, but including a negative pressure term due to
particle production is obtained from the {\it ansatz} $P_c=-\lambda$, where $\lambda$ is a positive constant 
having the dimension of an energy density. This {\it ansatz} is equivalent, as we shall see below, to the 
usual assumption that the particle production rate is proportional to the Hubble parameter if
the created particles are ``cold" and interact ``weakly". Consequently,
the enthalpy density is simply $h=\rho$ and the relation between the particle and the energy densities is given by
$\rho=nmc^2$, where $m$ is the rest mass of the created particles (supposed to be uncharged). Under these
conditions, eq.~\ref{particlerate} can be recast as
\begin{equation}
\frac{dn}{dt}+3Hn = \frac{3\lambda}{mc^2}H
\label{continuity}
\end{equation}
Notice that now the particle conservation equation has a (positive) source term proportional to the Hubble
parameter as mentioned above. Using the scale parameter $a$ as the independent variable 
instead of time, the equation above
can be rewritten as
\begin{equation}
\frac{dn}{d\lg a}+3n = \frac{3\lambda}{mc^2}
\label{density}
\end{equation}
and similarly for the energy density
\begin{equation}
\frac{d\rho}{d\lg a}+3\rho = 3\lambda
\label{energydensity}
\end{equation}
Integration of eq.~\ref{energydensity} is trivial and is given by
\begin{equation}
\rho = \lambda + \frac{(\rho_*-\lambda)}{a^3}
\end{equation}
where $\rho_*$ is the energy density when $a=1$ (the present time). The equation above can be recast
as
\begin{equation}
\rho = \rho_*\left(\Omega_v + \Omega_ma^{-3}\right)
\end{equation}
which is formally identical to the expression of the energy density in the case of the standard $\Lambda$CDM model,
if one identifies $\Omega_v = \lambda/\rho_*$ and $\Omega_m = 1-\Omega_v$. However, in the present case $\Omega_v$ is not
the density parameter associated to the cosmological term but to the creation rate. Similarly, $\Omega_m$ in the
present case is not equivalent to the matter parameter density as in the standard model, since  here the present
matter density is equal to the critical density. Integration of eq.~\ref{density} is also
trivial and gives for the evolution of the particle number density
\begin{equation}
n = n_*\left(\Omega_v + \Omega_ma^{-3}\right)
\end{equation}
where $n_*$ is the present particle number density.
For high redshifts or $a << 1$, the particle number density satisfies the condition $na^3 = n_*\Omega_m = constant$,
similar to the standard model. However, in the future, when $a >> 1$, the standard model predicts that either
for baryons or for dark matter, the particle
number density goes to zero due to the expansion of the universe while the present model predicts a constant density
equal to $n = n_*\Omega_v$ for dark matter particles. In other words, the present model predicts 
a future state of the universe constituted
by dark matter only since the density ratio $n_b/n$ goes to zero. This future behaviour of dark matter is
identical to that of the steady-state cosmology, since when $a >> 1$ the creation rate will be equal to
the expansion rate. It worth mentioning that the present matter density in this model is higher than
the standard $\Lambda$CDM by a factor of $1/\Omega_m$, a fact with observable consequences, as we shall see later. 
In reality, these are not the only differences with
the standard model. Since the ratio $s/n$ remains constant during the expansion, the evolution of the entropy
density is given by 
\begin{equation}
s = s_*\left(\Omega_v + \Omega_ma^{-3}\right)
\end{equation}
where $s_*$ is the present entropy density of dark matter. The equation above says that the evolution of the 
entropy density also differs from the standard model, with a temporal behaviour similar to that of the particle
density or, in other words, in the future the entropy of the universe will essentially dominated by that of dark matter.
Notice also that in the standard model, the entropy in a proper volume corresponding to a unit comoving 
volume ($sa^3$) is constant but in the present model there is entropy production at a rate
\begin{equation}
\frac{d(sa^3)}{dt} = 3s_*\Omega_va^3H
\end{equation}
Once the past behaviour of the present model is similar to that of the standard model, both models satisfy the
different tests (supernova distances, baryon acoustic peak, ``shift" parameter associated to the CMB, age
of the universe) if the parameters of both models are conveniently interpreted (see, for instance, reference \cite{lim}).

The particular cosmology here considered has a deceleration parameter $q$ formally similar to the standard
model, i.e.,
\begin{equation}
q = -\frac{\ddot{a}a}{\dot a^2} = \frac{1}{2}\frac{(\Omega_ma^{-3}-2\Omega_v)}{(\Omega_v+\Omega_ma^{-3})}
\end{equation}
Thus, the present value of this parameter is simply
\begin{equation}
q_* = \frac{1}{2}\left(1-3\Omega_v\right)
\end{equation}
Since observations of supernova distances indicate $\Omega_v = 0.7\pm 0.1$ (\cite{perl}), from the equation above
it results that $q_* = -0.55\pm 0.15$.

\section{The evolution of density perturbations}

In the present study, in order to obtain the linearized equations, the Newtonian approximation will be used.
This approximation is justified in the weak field limit, i.e., when the velocity of peculiar motions satisfies
the condition $V_p << c$ and the scale of the perturbations is much less than the Hubble radius 
(see ref. \cite{jim}). Notice that in the case of the ``standard" model, a fully relativistic treatment
leads essentially to same results obtained through the weak field approximation.

The first equation is the particle number conservation (eq.~\ref{continuity}), which can be recast as
\begin{equation}
\frac{dn}{dt} + \vec\nabla\cdot(n\vec U) = \alpha H
\label{continuity2}
\end{equation}
where $\alpha = 3\lambda/mc^2 = 3\Omega_vn_*$. The velocity field $\vec U$ includes the Hubble flow 
and peculiar velocities
resulting from gravitational interactions between density peaks. The second
equation expresses the gravitational aceleration in terms of the mass distribution, i.e.,
\begin{equation}
\vec\nabla\cdot\vec g = 4\pi Gmn
\label{poisson}
\end{equation}
Finally, the equation of motion, which in absence of pressure gradient terms can be simply written as
\begin{equation}
\frac{\partial\vec U}{\partial t} + \vec U\cdot\frac{\partial\vec U}{\partial r} + \vec g = 0
\label{motion}
\end{equation}
These equations are similar to those in the usual Newtonian approximation except by the fact
that now a source term is present in the continuity equation. Defining the perturbed quantities $n_1$, $\vec g_1$
and $\vec V_p$ by the relations,
\begin{eqnarray}
n(\vec r,t) = n_0(t) + n_1(\vec r,t)\\
\vec g(\vec r,t) = \vec g_0(\vec r,t) + \vec g_1(\vec r,t)\\
\vec U(\vec r,t) = H\vec r + \vec V_p(\vec r,t)
\end{eqnarray}
introducing the density contrast $\delta = n_1/n_0$, replacing the perturbed quantities into
eqs.~\ref{continuity2}, \ref{poisson}, \ref{motion}, expanding and retaining only first order terms, one obtains
respectively for the linearized continuity, force and motion equations 
\begin{equation}
\dot\delta + \vec\nabla\cdot\vec V_p = -\frac{4\pi Gm}{3H_0}\alpha\delta
\label{continuity3}
\end{equation}
\begin{equation}
\vec\nabla\cdot\vec g_1 = 4\pi Gmn_0\delta
\label{poisson2}
\end{equation}
and
\begin{equation}
\frac{\partial\vec V_p}{\partial t} + H_0\vec V_p + \vec g_1 = 0
\label{motion2}
\end{equation}
It should be emphasized that here the subscript $``0"$ denotes unperturbed values while quantities taken at present
time ($z=0$) are denoted by the subscript $``*"$. Writing the perturbed quantities in terms of the Fourier 
components, the system of linear equations above becomes
\begin{equation}
\dot\delta_k + \frac{i\vec k\cdot\vec V_k}{a} = -\frac{4\pi Gm\alpha}{3H_0}\delta_k
\label{continuity4}
\end{equation}
\begin{equation}
\frac{i\vec k\cdot\vec g_k}{a} = 4\pi Gmn_0\delta_k
\label{poisson3}
\end{equation}
\begin{equation}
\frac{\partial\vec V_k}{\partial t} + H_0\vec V_k + \vec g_k = 0
\label{motion3}
\end{equation}
After some lengthy but trivial algebra with these equations, one obtains finally for the evolution 
of the density contrast in the linear approximation
\begin{equation}
\ddot\delta_k + \left(2H_0 + \frac{4\pi Gm\alpha}{3H_0}\right)\dot\delta_k + 
4\pi Gm\left[\alpha\left(1-\frac{\ddot aa}{3\dot a^2}\right)-n_0\right]\delta_k = 0
\label{contrast}
\end{equation}
Notice if there is no particle creation ($\alpha = 0$) the usual equation describing the
evolution of the density contrast in absence of pressure gradients is 
recovered (see, for instance \cite{peebles}).

In order to solve numerically eq.~\ref{contrast}, it is convenient to introduce a dimensionless time
variable $\tau = t/t_*$, where $t_*$ is the present age of the universe. Using now the notation 
$\partial\delta_k/\partial\tau = \delta_k^{'}$ and $\partial^2\delta_k/\partial\tau^2 = \delta_k^{''}$,
eq.~\ref{contrast} can be written in a general form
\begin{equation}
\delta_k^{''}(\tau)+g_1(\tau)\delta_k^{'}(\tau)+g_2(\tau)\delta_k(\tau) = 0
\label{contrast2}
\end{equation}
where explicit expressions for the functions $g_1(\tau)$ and $g_2(\tau)$ depend on the adopted
cosmological model. For the standard $\Lambda$CDM model, these functions are given respectively by
\begin{equation}
g_1(\tau) = 2\beta_*f(a(\tau))=2\beta_*\sqrt{1+\chi a^{-3}(\tau)}
\end{equation}
and
\begin{equation}
g_2(\tau)=-\frac{3}{2}\beta_*^2\chi a^{-3}(\tau)
\end{equation}
In these equations, $\beta_* = H_*t_*\sqrt{\Omega_v}$, $\chi = \Omega_m/\Omega_v$ 
and the scale factor as a function of the dimensionless time is 
\begin{equation}
a(\tau) = \chi^{1/3}Sinh^{2/3}(3\beta_*\tau/2)
\label{scalea}
\end{equation}
When particle creation is included ($\alpha \not= 0$), the functions $g_1(\tau)$ and $g_2(\tau)$ 
become
\begin{equation}
g_1(\tau)=2\beta_*f(a(\tau))\left[1+\frac{3}{4}\frac{1}{f^2(a(\tau))}\right]
\end{equation}
and
\begin{equation}
g_2(\tau)=\frac{3}{2}\beta_*^2\left[\frac{(4+7\chi a^{-3}(\tau))}{2(1+\chi a^{-3}(\tau))}-f^2(a(\tau))\right]
\end{equation}
In both cases the parameter $\beta_*$ can be estimated from eq.~\ref{scalea}, using the condition $a(1)=1$.
Thus, $\beta_* = 2ArcSinh(\chi^{-1/2})/3$, implying that the only free parameter in these 
equations is $\Omega_v$.
 
Numerical solutions of these equations were obtained by adopting initial conditions appropriate to the
decoupling between matter and radiation at $z \sim 1100$ and using the fact that at these high redshifts
both models are similar to the Einstein-de Sitter cosmology. Moreover, since there is no continuous production
of baryons, the transition from an ``opaque" to a ``transparent" universe should occur at the same redshift
as in the canonical model. Thus, the evolution of the
density contrast when $a << 1$ is $\delta (t) \propto a(t)$. 

Our results indicate that for models with $\alpha \neq 0$, i.e., including particle creation,
the density contrast increases, reaches a maximum before the present time and then decreases if 
the parameter $\Omega_v$ is higher than a critical value equal to
 $\Omega_{v,crit}=0.666$. For this particular value, the maximum occurs at the present time and
for lower values, the maximum occurs in the future, i.e., when $a > 1$.
This behaviour is illustrated in fig. 1 where the evolution of the density contrast 
as a function of the scale factor and for different values of the parameter $\Omega_v$ is shown. For
$a < 0.03$ or $z > 35$, as expected, all models coincide since they follow the same behaviour of the
Einstein-de Sitter solution. 

\begin{figure*}
\includegraphics[width=1.4\textwidth]{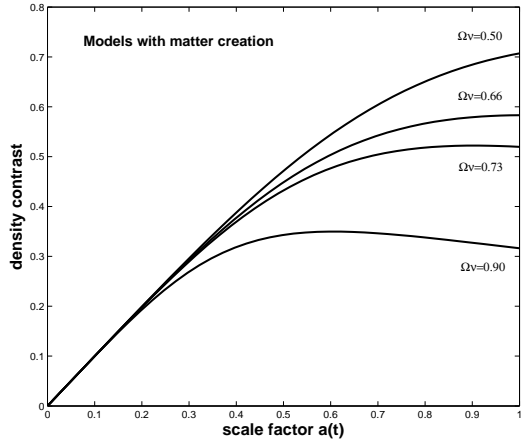}
\caption{Density contrast evolution as a function of the scale parameter. Labels
indicate different values of the parameter $\Omega_v$}
\end{figure*}
The existence of a maximum in the evolution of the density contrast is a consequence
of a fundamental difference between models including particle creation and the standard $\Lambda$CDM
cosmology. These differences are clearly seen in a close inspection of the coefficients appearing
in eq.~\ref{contrast} either when $\alpha = 0$ ($\Lambda$CDM model) or when $\alpha \neq 0$, particle
creation model. The first coefficient, corresponding to the second term on the left (or the coefficient
of the first derivative of the density contrast) is simply equal to $2H=2\dot a/a$ in the case of the
standard model. This is a damping mechanism related to the expansion of the universe. When particle
creation is included, this damping coefficient increases from $2H$ up to $2.36H$ for $a=1$ and $\Omega_v=0.73$.
Such a variation is a small effect not responsible for the existence of a maximum, although it contributes
to decrease slightly the amplitude of the density contrast. The second coefficient, corresponding to the
third term on the left of eq.~\ref{contrast}, can be recast as
\begin{equation}
4\pi Gm\left[\alpha\left(1+\frac{q}{3}\right)-n_0\right]
\end{equation}
where $q$ is the deceleration parameter. In the standard model ($\alpha=0$) the coefficient above
is negative, representing the well known gravitational instability. When particle production
is included, an extra term counterbalancing gravity appears and the amplitude of the density
contrast increases as far as the condition
\begin{equation}
\frac{2(1+\chi a^{-3})^2}{(4+7\chi a^{-3})}>1
\end{equation}
is satisfied which, for adequate values of $\chi = \Omega_m/\Omega_v$, is equivalent to the condition $a < 0.5$. 
Consequently,
depending on the value of $\Omega_v$, a maximum amplitude is attained in the late 
evolution of the density contrast, whose value depends on the particle creation rate. 

\begin{figure*}
\includegraphics[width=1.4\textwidth]{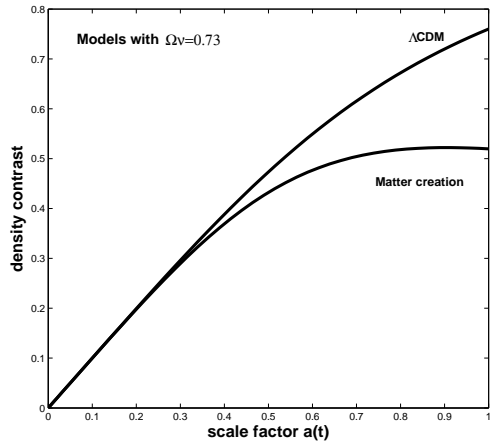}
\caption{Comparison between the density contrast evolution resulting from the standard
$\Lambda$CDM model and the present cosmology including matter creation.}
\end{figure*}
In fig. 2, the density contrast evolution for $\Omega_v = 0.73$ is shown for both the standard 
and the particle creation model. The density contrast of the later at the present time is about 30\% lower
than that derived for the standard model, with the maximum occurring at $z\sim 0.2$.

The effective damping of the density contrast induced by particle production poses some difficulties
to the process of galaxy and large-scale structure formation by gravitational instability. However,
other problems related with the growth of the density contrast exist in this model. For 
instance, the present root mean square value of the peculiar velocity of galaxies in the linear theory when
particle creation is absent is given by \cite{peebles}
\begin{equation}
<V^2_p>^{1/2} = Ha\left(\frac{dlg\delta(a)}{dlg a}\right)\sqrt{J_2}
\end{equation}
where $J_2 = 150h^{-2} Mpc^{2}$ is the second moment of the two-point correlation function of galaxies
\cite{davis}. The introduction of the particle creation process modifies the continuity equation, as we 
have seen previously, since a source term is now present. Consequently, the resulting equation
for the root mean square velocity of galaxies in the linear theory is now given by
\begin{equation}
<V^2_p>^{1/2}=Hf(a)a\left[1+\frac{3}{2}\frac{n_*}{n_0}\frac{\Omega_v}{f(a)}\right]\sqrt{J_2}
\end{equation}
where we have defined as in reference \cite{peebles} $f(a)=d\lg\delta/d\lg a$. Numerical solutions of
the equation above indicate that the present root mean square velocity practically independs on the parameter
$\Omega_v$, being equal to $\simeq 1200 km/s$. This is considerably higher than the value derived from
observational data, i.e., $325 km/s$ \cite{bruno}.  Moreover,
as mentioned before, the present matter density is greater than that of the standard $\Lambda$CDM model
by a factor of $\Omega_m^{-1}$. As a consequence, the predicted density of clusters of galaxies above
a given mass is higher either than predictions of the standard model or observational data. For
clusters with masses higher than $10^{15} M_{\odot}$, data by Bahcall and Cen \cite{cen} indicate that
the density is about $7.2\times 10^{-8} Mpc^{-3}$ while from the creation model a density
of about $3.8\times 10^{-6} Mpc^{-3}$ is predicted.

\section{Conclusions}

Cosmological models described by ``open" thermodynamic systems, i.e., including particle
creation at the expense the gravitation field can reproduce formally the past evolution 
of matter, mimicking the presence of a cosmological constant in Friedman equations. This
class of models results from the assumption that entropy and particle production rates
are equal and that the negative pressure resulting from the particle creation process is a constant.
This last hypothesis is equivalent to the assumption of a production rate proportional to
the Hubble constant. It is worth mentioning that in the steady-state cosmology the creation rate
required to compensate the expansion and to maintain a constant particle density is 
$\dot r_{ss} = 3H_*n_*$, which should be compared with the rate of the present model, given
by $\dot r_{cre} = 3\Omega_vH_*n_*$. Thus, the present model has two asymptotic behaviours: in
earlier times behaves as the standard model or the Einstein-de Sitter solution and in late
times behaves as the steady-state cosmology, except by the fact that  
the baryon-to-dark matter particle density ratio goes to zero.

Difficulties appear for the creation model when the evolution of density fluctuations is 
regarded closely. In this model, the particle production mechanism mimics also a damping
mechanism that limits the growth of the density contrast, constraining
the process of galaxy and large-scale structure formation.
Moreover, the predicted root mean square peculiar velocity of galaxies in the linear
theory is about four times higher than observations.  
The creation model has also difficulties to explain the observed
density of massive clusters of galaxies, predicting presently ($z=0$) an excess of objects by a factor
of fifty with respect to observations, 
consequence of the fact that the present matter density in this cosmology is equal to the critical 
value. Nevertheless it is important to mention
that the present analysis is based on a linear approach and that the inclusion of non-linear
terms, affecting the late evolution of the density contrast may eventually change our 
conclusions. Cosmological simulations including particle creation will be the next step 
to investigate these non-linear effects, that will be able to
corroborate or not the present study.


%

\end{document}